\documentclass{elsarticle}
\usepackage{amssymb}
\usepackage{amsmath}
\usepackage{calc}
\usepackage{graphicx}
\usepackage{bm}
\usepackage{amsfonts}
\usepackage{color}
\usepackage{epsfig}

\begin{document}

\begin{frontmatter}

\title{Wireless adiabatic power transfer}

\author[Sofia]{A. A. Rangelov\corref{cor1}}
\ead{rangelov@phys.uni-sofia.bg}
\author[Weizmann]{H. Suchowski\corref{cor1}}
\author[Weizmann]{Y. Silberberg}
\author[Sofia]{N. V. Vitanov}

\cortext[cor1]{These authors contributed equally to this work}
\address[Sofia]{Department of Physics, Sofia University, James Bourchier 5 blvd., 1164 Sofia, Bulgaria}
\address[Weizmann]{Department of Physics of Complex System, Weizmann Institute of Science, Rehovot 76100, Israel}

\begin{abstract}
We propose a technique for efficient mid-range wireless power
transfer between two coils, by adapting the process of adiabatic
passage for a coherently driven two-state quantum system to the
realm of wireless energy transfer. The proposed technique is shown
to be robust to noise, resonant constraints, and other interferences
that exist in the neighborhood of the coils.
\end{abstract}

\begin{keyword}
Wireless energy transfer, Adiabatic passage, Robust and efficient power transfer, Coupled mode theory.

\PACS 05.45.Xt \sep 32.80.Xx \sep 84.32.Hh \sep 85.80.Jm

\end{keyword}
\end{frontmatter}

\section{Introduction \label{Sec-Intro}}


The search for wireless power transfer techniques is as old as the invention
of electricity. From Tesla, trough the vast technological development during
the 20th century till recent days, many proposals have been made and
implemented in this research field. Established techniques for wireless
energy transfer are known both in the near- and far-field coupling regimes.
Examples for the former can be found in resonant inductive electric
transformers \cite{Stanley}, optical waveguides \cite{Ramo} and cavity
couplers \cite{Haus}. In the far field, one can find the mechanism of
transferring electromagnetic power by beaming a light source to a receiver
which is converted to usable electrical energy \cite{Brown}. Although these
techniques enable sufficiently efficient energy transfer, they suffer either
from the short-range interaction in the near-field,  or from the requirement
of line of sight in the far-field approaches. Recently, it was shown that
weakly radiative wireless energy transfer between two identical classical
resonant coils  is possible with sufficiently high efficiency \cite%
{Kurs,Karalis,Hamam}. This breakthrough was made possible by the application
of the coupled mode theory into the realm of power transfer. In this
experiment, Kurs \emph{et. al.} \cite{Kurs} showed that energy can be
transferred wirelessly at distances of about 2 meters (mid-range) with
efficiency of about 40\%.

Currently, most efficient wireless energy transfer devices rely upon  the
constraint of exact resonance between the frequencies of the emitter (the
source) and the receiver (the device, or the drain) coils \cite%
{Kurs,Karalis,Hamam}. When the frequency of the source is shifted from the
frequency of the device, due to lack of similarity between the coils or by
random noise (introduced, for example, by external objects placed close to
either coils),  a significant reduction of the transfer efficiency would
occur. In such case, one may implement a feedback circuit, as suggested in
Ref. \cite{Kurs}, in order to correct the reduction of the transfer
efficiency.

In this paper, we suggest a different approach to resolve the issues of the
resonant energy transfer process. Here,  we present a novel technique for
robust and efficient mid-range wireless power transfer between two coils,
by adapting the process of adiabatic passage (AP) for a coherently driven
two-state quantum system \cite{All87,Vit01a},  as will be explained in the
following sections. The adiabatic technique promises to be both efficient
and robust for variations of the parameters driving the process,  such as
the resonant frequencies of the coils and the coupling coefficient between
them.


\section{Overview of the coupled mode theory \label{two-state atom analogue}}


\begin{figure}[htb]
\begin{center}
\includegraphics[angle=0,width=0.8\columnwidth]{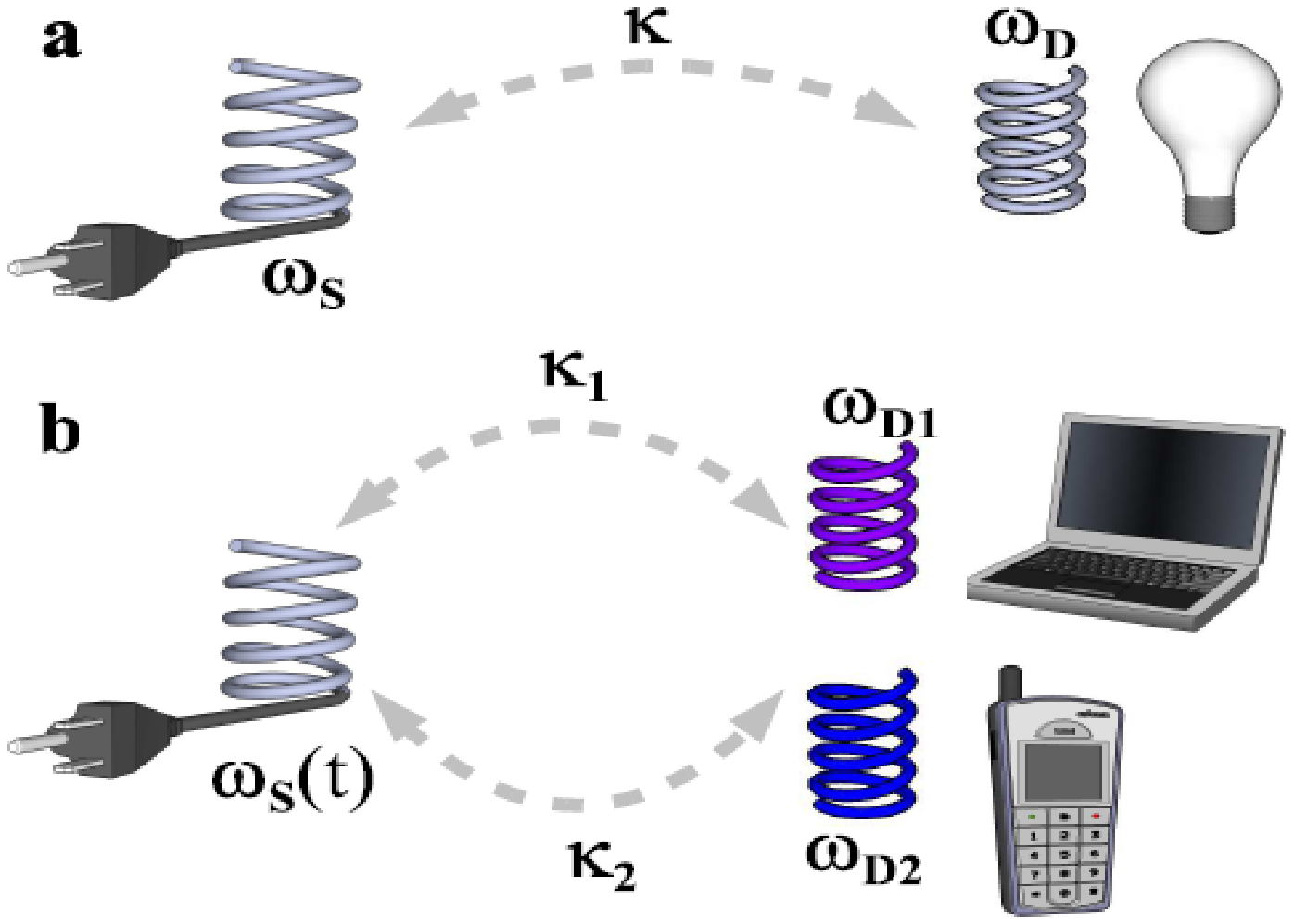}
\end{center}
\caption{(Color online) Methods for wireless energy transfer. (a) For
constant coil frequencies, efficient transfer from the source to a device
occurs only when $\protect\omega_{S}=\protect\omega_{D}$. (b) The proposed
adiabatic technique, with time-varying coil frequency $\protect\omega_S(t)$,
can transfer energy to multiple devices even when most of the time $\protect%
\omega_{S}(t) \neq \protect\omega_{D_1},\protect\omega_{D_2}$. }
\label{wireless dynamics}
\end{figure}

We follow the description of the coupled mode theory in the context of
wireless energy transfer as described in detail by Kurs \emph{et. al.} \cite%
{Kurs}. The interaction between two coils, in the strong-coupling regime, is
described by the coupled-mode theory, through the following set of two
differential equations \cite{Haus,Kurs}: \newline
\begin{equation}  \label{wireless equations}
i\frac{\,\text{d}}{\,\text{d}t}\left[
\begin{array}{c}
a_{S}(t) \\
a_{D}(t)%
\end{array}%
\right] = \left[%
\begin{array}{cc}
\omega_{S}(t) -i\Gamma_{S} & \kappa (t) \\
\kappa (t) & \omega_{D}(t) -i\Gamma_{D}-i\Gamma_{W}%
\end{array}%
\right] \left[%
\begin{array}{c}
a_{S}(t) \\
a_{D}(t)%
\end{array}%
\right] .
\end{equation}
Here, $a_{S}(t) $ and $a_{D}(t)$ are defined so that the energies contained
in the source and the drain are, respectively, $\vert a_{S}(t)\vert^2$ and $%
\vert a_{D}(t) \vert^2$. $\Gamma_{S}$ and $\Gamma_{D}$\ are the intrinsic
loss rates (due to absorption and radiation) of the source and the drain
coils, respectively, and the extraction of work from the device is described
by the term $\Gamma_{W}$. The intrinsic frequencies of the source and drain
coils are $\omega_{S}(t)$ and $\omega_{D}(t)$; these are given explicitly as
\begin{equation}
\omega_m(t) =1/\sqrt{L_m(t) C_m(t) }\qquad (m=S,D),
\end{equation}
where $L_{S,D}(t)$ and $C_{S,D}(t)$ are the inductance and the capacitance,
respectively, of the source and the drain coils. The coupling coefficient
between the two coils reads 
\begin{equation}
\kappa (t) = M(t) \sqrt{\frac{\omega_S(t) \omega_D(t) }{L_S(t) L_D(t)}},
\label{coupling}
\end{equation}
where $M(t)$ is the mutual inductance of the two coils. The source coil is a
part of the driving circuit and is periodically recharged, while the energy
is transferred wirelessly to the device coil. The dynamics of such process
in the case of static (time independent) resonance frequencies, as describe
in Ref. \cite{Kurs} is illustrated in Fig. \ref{wireless dynamics} (top).

The evolution of Eq. (\ref{wireless equations}) is connected to the dynamics
of the Schr\"{o}dinger equation for a two-state atom written in the
rotating-wave approximation \cite{All87,Vit01a}. The variables $a_S(t)$ and $%
a_D(t)$ can be identified as the probability amplitudes for the ground state
(corresponding to the source) and the excited state (corresponding to the
drain), respectively. The coupling between the coils are analogous to the
coupling coefficient of the two-state atom (also known as the Rabi
frequency), which is proportional to the atomic transition dipole moment $%
\mathbf{d}_{12}$ and the laser electric field amplitude $\mathcal{E}(t) $:  $%
\Omega (t) = -\mathbf{d}_{12}\cdot \mathcal{E}(t)/\hbar$ \cite{All87,Vit01a}%
. The difference between the resonant frequencies of the two coils
corresponds to the detuning $\Delta (t)$ in the two-state atom:  $\Delta (t)
=\omega_{D}(t) -\omega_{S}(t)$.

The power transfer method was demonstrated for the resonant case of $%
\omega_{S} = \omega_{D} = \text{const}$, which is the case of $\Delta=0$ in
atomic physics. However, the power transmitted between the coils drops
sharply as the system is detuned from resonance, i.e. for the case of $%
\omega_{S} \neq \omega_{D}$. Also, any time dependent dynamics or change of
coupling strengths between the coils can results in lower energy transfer
between the coils.


\section{Adiabatic wireless energy transfer \label{AP formalism}}


In the following, we develop a systematic framework of the adiabatic
criteria in the context of wireless energy transfer. The technique of
adiabatic passage was successfully implemented in other research fields,
such as nuclear magnetic resonance (NMR) \cite{NMR1,NMR2}, interaction of
coherent light with two level atoms \cite{All87,Vit01a}, or in sum frequency
conversion techniques in nonlinear optics \cite{Suchowski08,Suchowski09}.
This dynamical solution requires a time-dependent intrinsic frequency change
of the source coil $\omega_{S}(t)$. The variation of the frequency  should
be adiabatic (very slow) compared to the internal dynamics of the system
that is determined by the coupling coefficient.

We will first assume that the loss rates $\Gamma_{S}$, $\Gamma_{D}$ and $%
\Gamma_{W}$ are zero  and write Eq. (\ref{wireless equations}) in the
so-called \emph{adiabatic basis} \cite{All87,Vit01a}  (for the two-state
atom this is the basis of the instantaneous eigenstates of the Hamiltonian):
\begin{equation}  \label{wireless equations in adiabatic basis}
i\frac{\,\text{d}}{\,\text{d}t} \left[
\begin{array}{c}
b_-(t) \\
b_+(t)%
\end{array}%
\right] =\left[
\begin{array}{cc}
-\varepsilon (t) & i\dot{\vartheta}(t) \\
i\dot{\vartheta}(t) & \varepsilon (t)%
\end{array}%
\right] \left[
\begin{array}{c}
b_-(t) \\
b_+(t)%
\end{array}%
\right] ,
\end{equation}
where the dot denotes a time derivative and
\begin{subequations}
\begin{align}
\varepsilon(t) &= \tfrac12\sqrt{4\kappa(t)^2 +\Delta(t)^2 }, \\
\tan 2\vartheta (t) &= \frac{2\kappa (t) }{\Delta (t) }\,.  \label{tita}
\end{align}
The connection between the original amplitudes $a_{S}(t) $ and $a_{D}(t) $
and the adiabatic ones $b_-(t) $ and $b_+(t)$ is given by
\end{subequations}
\begin{subequations}
\label{adiabatic states}
\begin{align}
b_-(t)&= a_{S}(t) \cos \vartheta (t)-a_{D}(t) \sin\vartheta (t), \\
b_+(t)&= a_{S}(t) \sin \vartheta (t)+a_{D}(t) \cos\vartheta (t).
\end{align}
When the evolution of the system is adiabatic,  $\vert b_-(t)\vert^2$ and $%
\vert b_+(t)\vert^2$ remain constant \cite{Messiah}. Mathematically,
adiabatic evolution means that the non-diagonal terms in Eq. \eqref{wireless
equations in adiabatic basis} are small compared to the diagonal terms and
can be neglected. This restriction amounts to the following \emph{adiabatic
condition} on the process parameters \cite{All87,Vit01a}:
\end{subequations}
\begin{equation}  \label{adiabatic condition}
\vert \dot{\kappa}(t) \Delta (t) -\kappa (t)\dot{\Delta}(t)\vert \ll
[4\kappa(t)^2 +\Delta(t)^2]^{3/2}.
\end{equation}
Hence adiabatic evolution requires a smooth time dependence of the coupling $%
\kappa (t)$ and the detuning $\Delta (t)$,  long interaction time, and large
coupling and/or large detuning. In the adiabatic regime $\vert
b_{\pm}(t)\vert^2 = \text{const}$,  but the energy contained in the source
and the drain coil $\vert a_{S,D}(t)\vert^2$ will \emph{vary} if the mixing
angle $\vartheta (t)$ varies;  thus adiabatic evolution can produce energy
transfer between the two coils.

If the detuning $\Delta (t) $ sweeps slowly from some large negative value
to some large positive value (or vice versa), then the mixing angle $%
\vartheta(t)$ changes from $\pi /2$ to $0$ (or vice versa). With the energy
initially in the first coil, the system will stay adiabatically in $b_-(t)$\
and thus the energy will end up in the second coil. Therefore the detuning
sweep (i.e. the frequency chirp) will produce complete energy transfer.
Furthermore, AP is not restricted to the shape of the coupling $\kappa (t)$
and the detuning $\Delta (t)$  as far as the condition \eqref{adiabatic
condition} is fulfilled and the mixing angle $\vartheta (t)$ changes from $%
\pi/2$ to 0 (or vice versa).

The variation of the detuning $\Delta(t)$ can be achieved by changing the
capacitance (or the inductance) of one, or the two coils. The time variation
of the coupling $\kappa (t)$ can be achieved,  for example, with the
rotation of one coil (or two coils), thereby changing the geometry and thus
the mutual inductance $M(t)$ of the two coils.

\begin{figure}[htb]
\begin{center}
\includegraphics[angle=0,width=100mm]{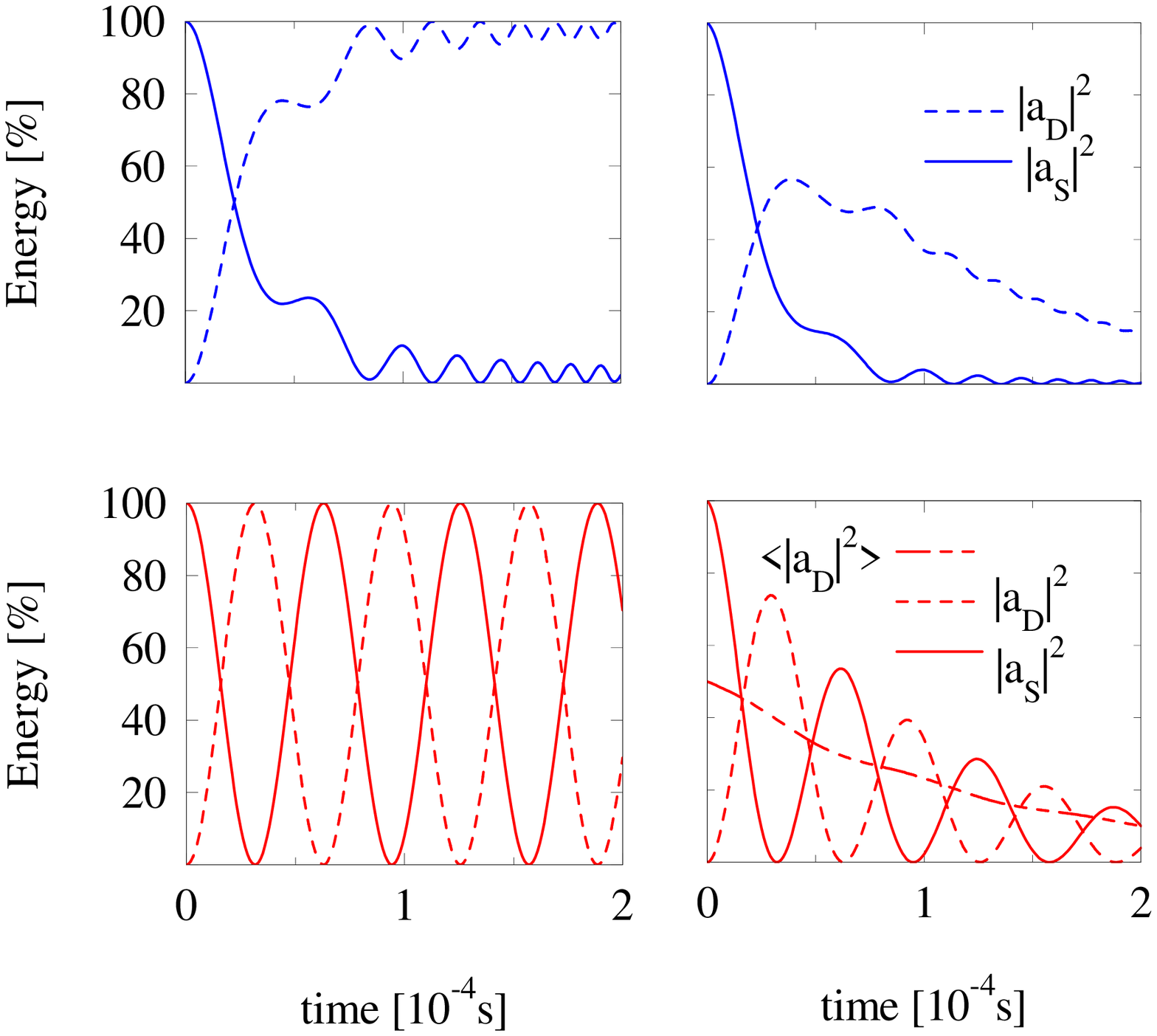}
\end{center}
\caption{(Color online) Comparison between the energy transfer as a function
of time for the AP case (top frames) and the static case (bottom frames),
with no losses (left frames) and with loss $\Gamma _{S,D}=2\times 10^{3}$ s$%
^{-1}$ (right frames). In all the graphs, solid line refers to the source
coil, and dashed line refers to the device coil. For the static case, the
functions $\protect\kappa (t)$ and $\Delta (t)$ are given by Eqs.~
\eqref{resonance} with the following parameters: $\protect\kappa %
_{0}=4\times 10^{4}$ s$^{-1}$, $\protect\delta =0$ s$^{-1}$, whereas for the
AP case, they follow Eqs.~\eqref{AP} with the following parameters: $\protect%
\kappa _{0}=4\times 10^{4}$ s$^{-1}$, $\protect\delta =2\times 10^{5}$ s$%
^{-1}$, $\protect\beta =3\times 10^{9}$ s$^{-2}$.}
\label{fig2}
\end{figure}

When the loss rates are nonzero, the dynamics become more complicated and
more realistic. Nevertheless, the essence of AP remains largely intact, if
one follow another important constraint,
\begin{equation}
\Gamma _{S,D}<\kappa _{0}<|\Delta (t)|.
\end{equation}%
which states that the coupling coefficient also should be larger than the
loss rates, and that the initial and final detunings are larger than the
coupling coefficient \cite{All87,Vit01a}. The physical reasoning behind it
is that the dynamics should be faster then the damping rates that exist in
the system (mainly on the device) and not only adiabatic. In Fig.~\ref{fig2}
we compare the resonant (static) and adiabatic mechanisms, without losses
(left frames) and with losses (right frames). For the numerics, we used the
following coupling and detuning for the resonant case:
\begin{subequations}
\label{resonance}
\begin{eqnarray}
\kappa (t) &=&\kappa _{0},\qquad  \\
\Delta (t) &=&\delta ,
\end{eqnarray}%
\end{subequations}
and for the adiabatic mechanism:
\begin{subequations}
\label{AP}
\begin{eqnarray}
\kappa (t) &=&\kappa _{0},\qquad  \\
\Delta (t) &=&\delta +\beta (t-t_{0}),
\end{eqnarray}%
\end{subequations}
where for our simulations we set $t_{0}=10^{-4}$ $s$. As can be seen in Fig. %
\ref{fig2} the energy in the static case oscillates back and forth between
the two coils. In AP, once the energy is transferred to the drain coil it
stays there. This feature of AP is used to minimize the energy losses from
the source coil. We see that when following the adiabatic constraints, the
AP process outperforms the static resonant method.

To describe the efficiency of the proposed technique we use the efficiency
coefficient $\eta $, which is the ratio between the work extracted from the
drain for the time interval $T$ divided by the total energy (absorbed and
radiated) for the same time interval,
\begin{equation}
\eta =\frac{\Gamma _{W}\int_{0}^{T}|a_{D}(t)|^{2}dt}{\Gamma
_{S}\int_{0}^{T}|a_{S}(t)|^{2}dt+(\Gamma _{D}+\Gamma
_{W})\int_{0}^{T}|a_{D}(t)|^{2}dt}.  \label{efficiency coefficient}
\end{equation}%
In the static steady state case, the efficiency reads \cite{Kurs}
\begin{equation}
\eta =\frac{\Gamma _{W}|a_{D}|^{2}}{\Gamma _{S}|a_{S}|^{2}+(\Gamma
_{D}+\Gamma _{W})|a_{D}|^{2}}.
\end{equation}

\begin{figure}[htb]
\begin{center}
\includegraphics[angle=0,width=100mm]{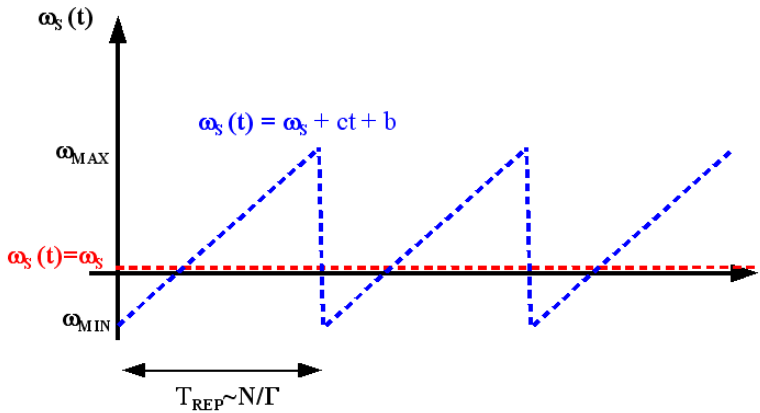}
\end{center}
\caption{(Color online) The function of the resonant frequency of the source
coil. The static case is shown in red, and the repeated linear case (which
is the simplest realization or the AP dynamics) is shown in blue.}
\label{fig3}
\end{figure}

As can be seen from Eq.~\eqref{efficiency coefficient}, in order to maximize
$\eta$, one should reduce the time that the energy stays in the source coil.
This cannot be obtained in the resonant (static) case, because the energy
oscillates back and forth between the source and the drain coils.
Nevertheless, in the AP mechanism, there is only one transition between the
source and the device, so it should be chosen to be as early as possible.
The AP dynamics (or any other time-dependent dynamics) should also repeat
itself after some repetition time $T_{rep}$. This is illustrated in Fig. \ref%
{fig3}. The time scale is of the order of several ``loss times'' (equal to $%
1/\Gamma_{S,D,W}$), in this way we ensure that each cycle of coil charging
begins, after the consumption of all the energy from the previous cycle,
otherwise interference appears, which will be difficult to be predicted
analytically. For the propose AP technique we assume that energy is
instantly loaded into the source coil without loss in the beginning of each
cycle.

Another important measurement is the amount of energy transferred from the
source coil to the device,  which is the useful energy consumed as a
function of time. This measurement is in fact equal to the nominator in Eq.~%
\eqref{efficiency coefficient}.

\begin{figure}[htb]
\begin{center}
\includegraphics[angle=0,width=80mm]{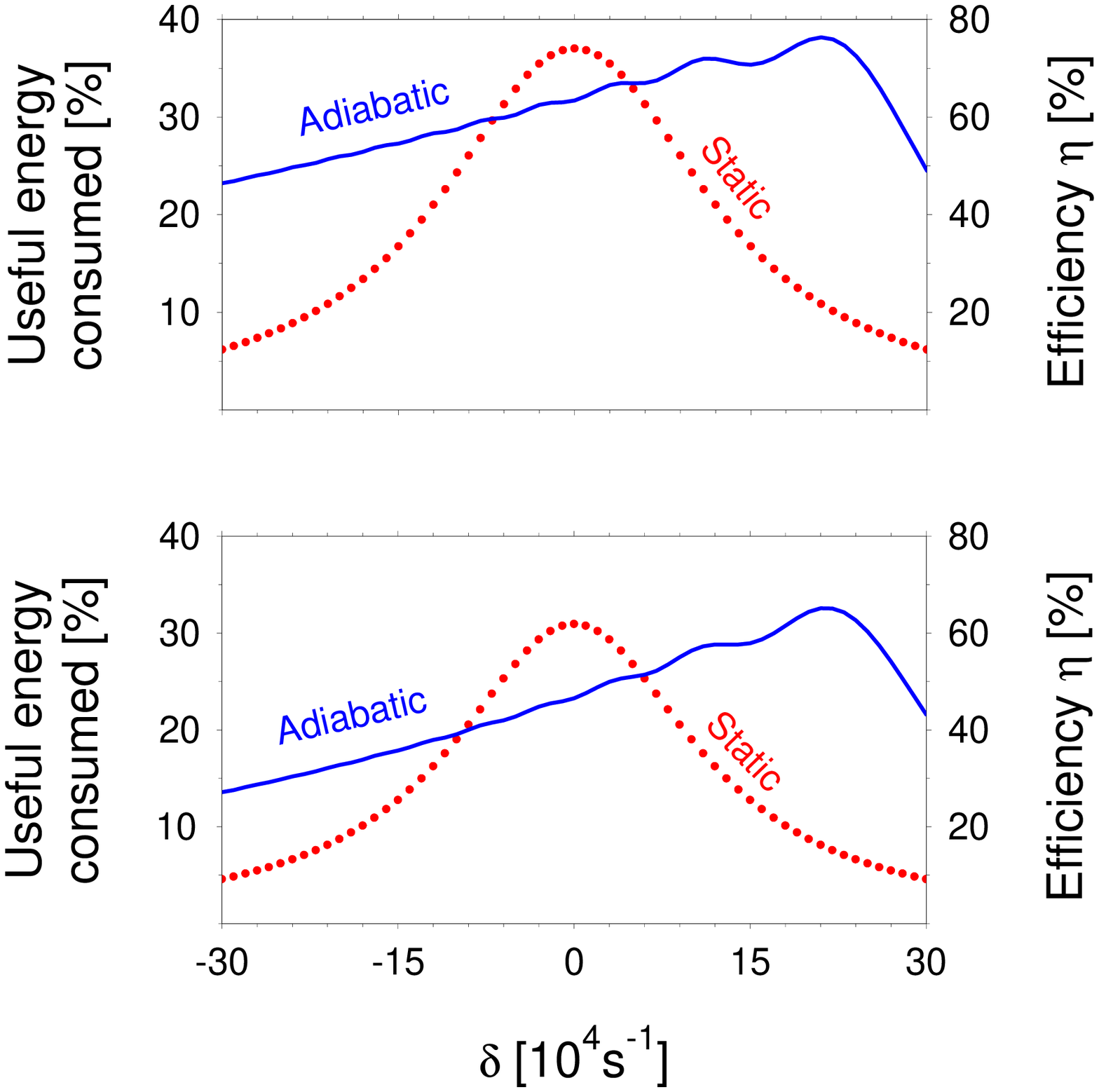}
\end{center}
\caption{(Color online) Efficiency coefficient $\protect\eta $
(right y-axes) and the useful energy consumed (left y-axes) as a
function of the static detuning. The blue solid line depicts AP
and the red dotted line is for the static method. The functions
$\protect\kappa (t)$ and $\Delta (t)$ are given by
Eqs.~\eqref{resonance} for the static method and \eqref{AP} for
AP, with the following parameter values: (top frames) $\protect\kappa %
_{0}=5\times 10^{4}$ s$^{-1}$, $\protect\beta =3\times 10^{9}$ s$^{-2}$, $%
\Gamma _{W}=10^{4}$ s$^{-1}$, $\Gamma _{S}=\Gamma _{D}=\protect\kappa %
_{0}/30=0.17\times 10^{4}$ s$^{-1}$. (bottom frames) $\protect\kappa %
_{0}=5\times 10^{4}$ s$^{-1}$, $\protect\beta =3\times 10^{9}$ s$^{-2}$, $%
\Gamma _{W}=10^{4}$ s$^{-1}$, $\Gamma _{S}=\Gamma _{D}=\protect\kappa %
_{0}/17=0.29\times 10^{4}$ s$^{-1}$. }
\label{fig4}
\end{figure}

\begin{figure}[htb]
\begin{center}
\includegraphics[angle=0,width=80mm]{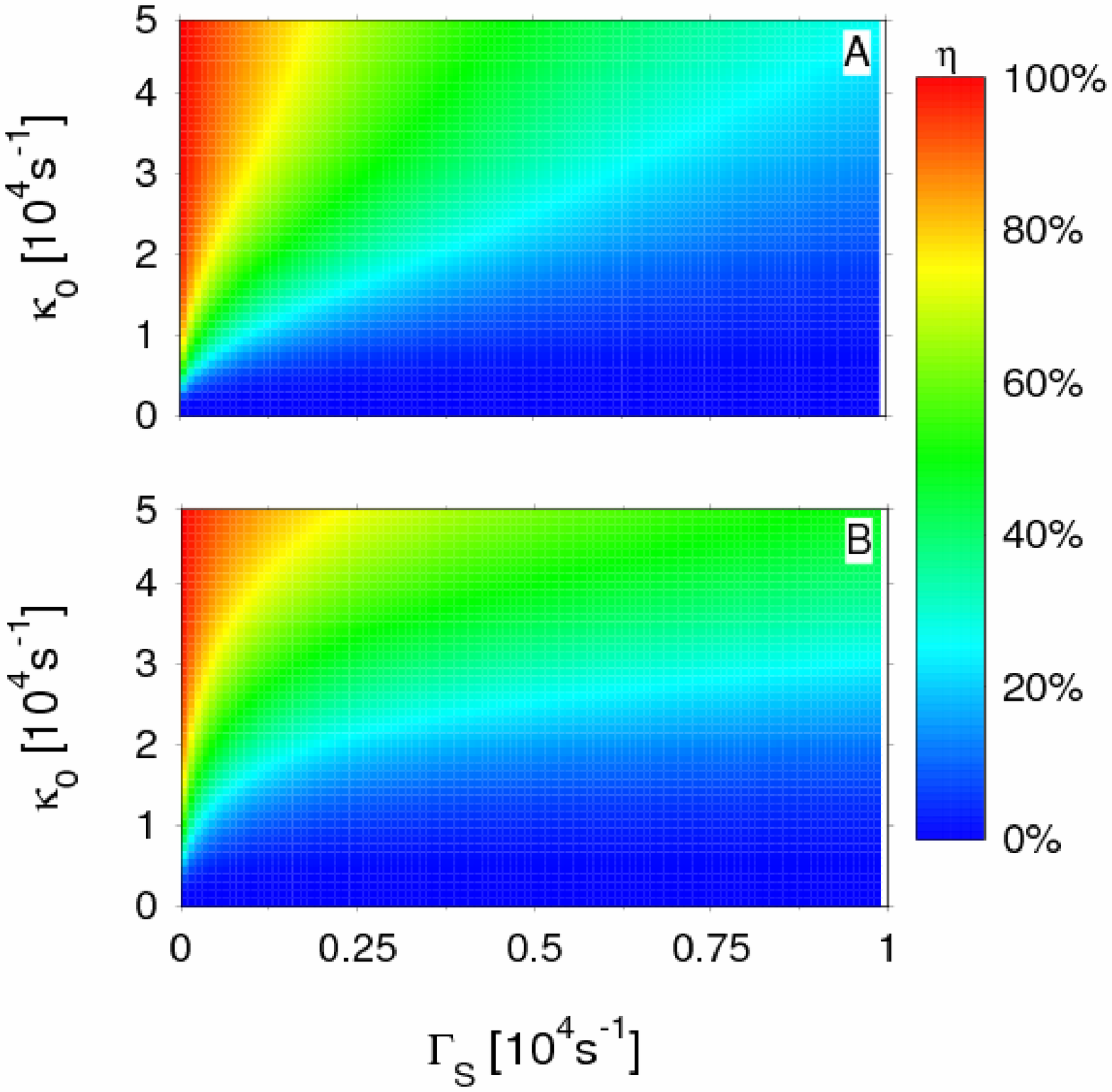}
\end{center}
\caption{(Color online) Efficiency coefficient $\protect\eta $ for
AP (top frame) and the static case (bottom frame) versus the loss
rate $\Gamma _{S}\equiv \Gamma _{D}$ and the coupling coefficient
$\protect\kappa _{0}$. The functions $\protect\kappa (t)$ and
$\Delta (t)$ are given by Eqs.~ \eqref{resonance} for the static
method and \eqref{AP} for AP, with the
following parameter values: $\protect\beta =3\times 10^{9}$ s$^{-2}$, $%
\protect\delta =2\times 10^{5}$ s$^{-1}$, $\Gamma_{W}=10^4$
s$^{-1}$. } \label{fig5}
\end{figure}

\section{Simulations\label{Simulations}}


In order to compare AP and the static mechanisms for energy transfer,
several sets of simulations were performed, measuring  both conversion
efficiency and total energy consumed by the device:

\begin{itemize}
\item Comparison between efficiencies as a function of the detuning,  for
different distances between the coils (determined by the ratio $%
\kappa/\Gamma_{S,D}$);

\item Influence of variations in the coupling and the loss coefficients;

\item Robustness of the adiabatic energy transfer, for time-dependent
coupling coefficient.
\end{itemize}

First, the effect of different resonant frequency between the source coil
and the device coil on the total wireless transfer efficiency were examined.
For that we changed only the resonance frequency of the device in each
numerical run, for two different distances between the coils: 0.8 meter and
1 meter between the coils (where we used the same units and notations as
reported in Ref. \cite{Kurs}). Fig. \ref{fig4} shows that AP is less
sensitive to the static detuning $\delta$. As can be shown, the maximal
conversion efficiency is achieved for the resonance approach is symmetric
about zero detuning, while AP is asymmetric with its maximal efficiency
value shifted toward the positive detuning. The explanation is that then the
energy transfer occurs at early stage and therefore the energy stays less
time in the source coil.

Next, we examine the effect of different coupling and loss coefficients.
Fig. \ref{fig5} shows contour plots of the efficiency coefficient $\eta$  as
a function of the coupling $\kappa_{0}$ and the loss rate of the source and
the drain coils (where we assumed $\Gamma_{S}=\Gamma_{D}$),  where we assume
fix values of $\Gamma_{W}$ (i.e. cannot be changed for different value of
coupling and loss coefficients). The upper frame presents results for the AP
technique, while the lower frame is for the static method. AP is obviously
more robust to the change in the parameter values compared to the static
method. The scheme proposed here, which does not require feedback control,
is therefore an alternative to the scheme suggested in \cite{KaralisPhD}.

We also wanted to check the effect of time-dependent coupling to the
dynamics of energy transfer, which is a more realistic modelling to the
process. When the detuning is varied between the two coils, the coupling
changes as well (can be inferred from Eq.~\eqref{coupling}). The maximal
coupling coefficient value is expected to be obtained when the detuning is
zero. For that, we chose the following time dependent detuning and coupling
coefficient, respectively:
\begin{subequations}
\label{shapes time dependent}
\begin{gather}
\Delta (t) =\delta + \beta (t-t_{0}) , \\
\kappa (t) = \kappa_{0}-\sqrt{|\Delta (t)| },
\end{gather}
where for our simulations we set $t_{0}=10^{-4}$ $s$. Fig. \ref{fig6} shows
the corresponding energy transfer efficiency as a function of time, along
with the detuning and coupling.
\begin{figure}[htb]
\begin{center}
\includegraphics[angle=0,width=80mm]{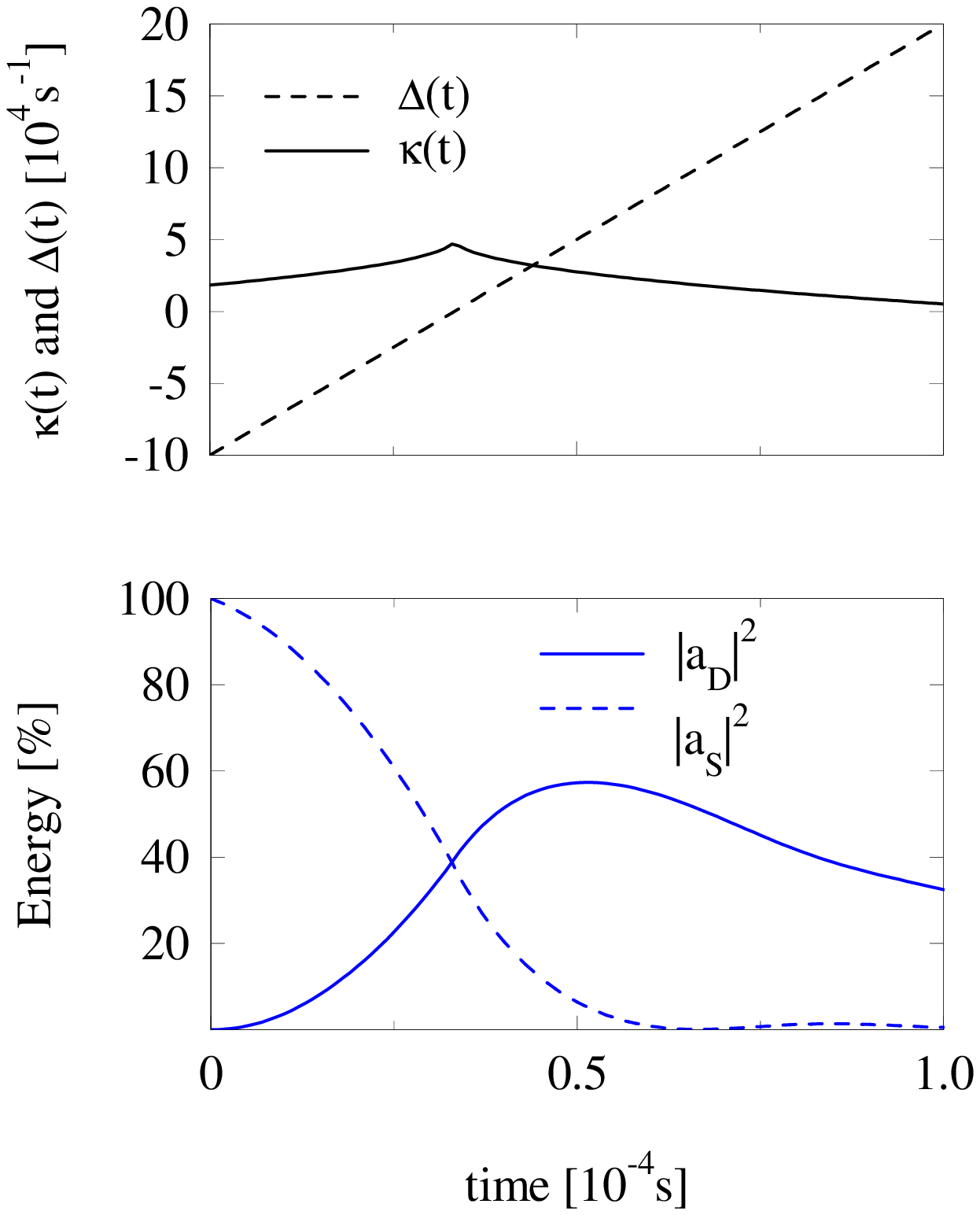}
\end{center}
\caption{(Color online) Coupling and detuning (top frame) and energy
transfer efficiency (lower frame) in AP for the case when the variation of
the detuning induces time variation in the coupling; the latter are given by
Eqs.~\eqref{shapes time dependent}. We have used the following parameters: $%
\protect\kappa_{0}=5\times 10^4$ s$^{-1}$, $\protect\beta=3\times 10^9$ s$%
^{-2}$, $\protect\delta=2\times 10^5$ s$^{-1}$, $\Gamma_{S}=\Gamma_{D}=%
\Gamma_{W}=3\times 10^3$ s$^{-1}$. }
\label{fig6}
\end{figure}


\section{Conclusions}


In conclusion, we have shown that the technique of adiabatic passage, which
is well known in quantum optics and nuclear magnetic resonance,  has analog
in the wireless energy transfer process between two circuites. The factor
that enables this analogy is the equivalence of the Schr\"{o}dinger equation
for two-state system, to the coupled-mode equation which describes the
interaction between two classical coils in the strong-coupling regime.

The proposed procedure transfers energy wirelessly in effective, robust
manner between two coils, without being sensitive to any resonant
constraints and noise compared to the resonant scheme demonstrated
previously. The application of this mechanism enables efficient energy
transfer to several devices as well as optimizing the transfer for several
distances and noise interferences.

This work has been supported by the European Commission projects EMALI and
FASTQUAST, the Bulgarian NSF grants D002-90/08 and DMU02-19/09 and Sofia
University Grant 074/2010.

\end{subequations}

\end{document}